\documentclass[10pt,aps,pre,superscriptaddress,floatfix,twocolumn,showpacs,amsmath,amssymb]{revtex4-1}
\usepackage[english]{babel}
\usepackage{booktabs}
\usepackage{natbib}
\usepackage{graphicx}
\usepackage{hyperref}

\newcommand{\ang}[1]{\left \langle #1 \right\rangle}
\newcommand{\abs}[1]{\left| #1 \right|}
\newcommand{\Ac}{\mathcal{A}}
\newcommand{\Lc}{\mathcal{L}}
\newcommand{\var}{\mathrm{Var}}

\begin{document}

\title{Minimal model of financial stylized facts}

\author{Danilo Delpini}
\email[Electronic address: ]{danilo.delpini@pv.infn.it}
\affiliation{Dipartimento di Economia Politica e Metodi Quantitativi,
Universit\`a degli Studi di Pavia}
\affiliation{INFN - Sezione di Pavia, via Bassi 6, Pavia, 27100, Italy}
\affiliation{CeRS - IUSS, V.le Lungo Ticino Sforza 56, Pavia, 27100, Italy}

\author{Giacomo Bormetti}
\email[Electronic address: ]{giacomo.bormetti@sns.it}
\affiliation{Scuola Normale Superiore, Piazza dei Cavalieri 7, Pisa, 56126, Italy}
\affiliation{INFN - Sezione di Pavia, via Bassi 6, Pavia, 27100, Italy}

\date{\today}

\begin{abstract}
In this work we afford the statistical characterization of a linear Stochastic Volatility Model
featuring Inverse Gamma stationary distribution for the instantaneous volatility. We detail the
derivation of the moments of the return distribution, revealing the role of the Inverse
Gamma law in the emergence of fat tails, and of the relevant correlation functions. We also propose
a systematic methodology for estimating the parameters, and we describe the empirical analysis of
the Standard \& Poor 500 index daily returns, confirming the ability of the model to capture many of
the established stylized fact as well as the scaling properties of empirical distributions over different
time horizons.
\end{abstract}


\maketitle

\section{Introduction}

A large number of empirical studies has shown that financial time series
exhibit statistical features strongly departing from the Gaussian behavior.
This finding dates back to the work of Mandelbrot \cite{Mandelbrot:1963}, whose
attention was mainly focused in recognizing realizations of stable processes,
and to the analysis of Fama~\cite{Fama:1965} concerning the long tailed nature
of the Dow Jones Industrial Average single components. Since these fundamental
contributions the modeling of financial returns has considerably grown, and
very heterogeneous models, able to reproduce the degree of asymmetry and the
excess of kurtosis of the measured distributions, have been proposed.
A non exhaustive list includes approaches developing from specific
distributional assumptions, as it is the case of the L\'evy
flights~\cite{Mantegna_book,PhysicaA.179.232,PhysRevLett.73.2946}, the
Generalized Student-$t$ or Tsallis
distributions~\cite{Bouchaud_book:2000,PhysRevLett.89.098701,Bormetti2007532}
and the exponential one~\cite{PhysicaA.329.178}.
Past empirical analysis have also proved the existence of non trivial scalings
of higher order correlations between returns at different times, pointing
toward the existence of a secondary stochastic process, as fundamental as that
of the price, governing the volatility of returns.
Many effective mechanisms allowing to reproduce the observed correlation
structures, where the stochastic nature of the volatility plays a central role,
were proposed. Discrete time models include ARCH-GARCH
processes~\cite{1982,Bollerslev1986307}, and multifractal
models~\cite{EurPhysJB.17.537,borland2005dynamics}, inspired by the cascades
originally introduced by Kolmogorov in the context of turbulent flows.  As far
as continuous time approaches are concerned, fractional Brownian motion and
stochastic volatility models have been extensively
analyzed. For a review of the latter approach we
suggest~\cite{Fouque_Papanicolau_Sircar_book} and remind the reader to the
discussion in Section~\ref{sec:model}. Focusing on the continuous time
stochastic volatility framework, in this work we aim at reproducing many of the
above mentioned facts which are generally accepted as universal evidences,
shared among different markets in different times.

The structure of the paper is the following. After introducing a general class of stochastic 
models driving the evolution of the volatility, in Section~\ref{sec:model} we concentrate on a 
linear one able to reproduce an Inverse Gamma distribution in the long run. 
In Section~\ref{sec:fatTails} we
detail the derivation of the moments of the probability density function $p(x;t)$ of the returns over
the time lag $t$, taking into account explicitly the time at which the $Y$
process has started and deriving rigorously the stationary limit of the
volatility.
We describe the mechanism through which the power law distribution of $\sigma$
induces fat tails on $p(x;t)$ for all the finite time lags.
In Sections~\ref{sec:lev} and~\ref{sec:autoc} we derive the analytical
expressions of the leverage correlation and the volatility autocorrelation
functions respectively.
In Section~\ref{sec:estimation} we propose a systematic methodology for
estimating the model parameters, and we apply it to the time series of the
daily returns of the Standard \& Poor 500 index. The relevant
conclusions, along with possible perspectives, will be summarized in
Section~\ref{sec:concl}.

\section{The Model}\label{sec:model}

We consider a model where the asset price 
\begin{equation*}
	S_t = S_0 \,\exp\left(\mu \,t + X_t\right)
\end{equation*}
is a function of the stochastic centred log-return $X_t$ and $\mu$ is a constant drift coefficient. 
We assume that $X_t$ can be modeled with
the following stochastic differential equation (SDE)
\begin{equation}
	dX_t = \sigma_t\,dW_{1,t}\,,
	\label{eq:dXeq}
\end{equation}
where $\sigma_t$ is the instantaneous volatility of the price and
$dW_{1,t}$ is the increment of a standard Wiener process. Since $X_0=0$,
from the above assumption we have that $\ang{X_t}=0$ and
$\ang{\ln(S_t/S_0)}=\mu t$ for all $t$. In the context of stochastic
volatility models (SVMs) the instantaneous volatility is assumed to be a
function of an underlying driving process $Y_t$, i.e. $\sigma_t=\sigma(Y_t)$. Typically,
the dynamics chosen for $Y_t$ corresponds to a particular case of the
following general multiplicative diffusion process
\begin{equation}
  dY_t = (a Y_t + b)\,dt+\sqrt{c \,Y_t^2 + d \,Y_t + e}\,dW_{2,t}\,,
  \label{eq:Y_MNDP}
\end{equation}
with suitable constraints on the parameters, in order to ensure the well
definiteness of the process. Moreover, the two standard Wiener processes
$W_{1,2}$ are possibly correlated
\begin{equation}
  \ang{dW_{1,t_1}\,dW_{2,t_2}} = \rho \,\delta(t_1-t_2) \,dt\,,
  \label{eq:Wcorrel}
\end{equation}
with $\rho\in[-1,1]$, which is necessary to account for skewness effects and
for the return-volatility correlation. For instance, in the Stein-Stein
model~\cite{RevFinancStudies.4.727,MP:2002} the volatility is linear,
$\sigma_t\propto Y_t$, and $Y_t$ follows a mean reverting Ornstein-Uhlenbeck
dynamics corresponding to $a<0$, $b>0$, $c=d=0$. Under the same $Y$ dynamics but
with $\sigma_t\propto \exp{(Y_t)}$ we obtain the exponential
Ornstein-Uhlenbeck model~\cite{JFinancQuantAnal.22.419,MP_Qfin:2006}.  In the Heston
model~\cite{Heston1993,DragulescuYakovenko2002} $\sigma_t=\sqrt{Y_t}$ and $Y_t$ evolves according
to a Cox-Ingersoll-Ross  dynamics, stemming from~\eqref{eq:Y_MNDP} by taking
$a<0$, $b>0$ with $c=e=0$. Finally, in the Hull-White model the volatility has 
the same functional dependence as in Heston, but $Y_t$ has a Log-Normal (non mean 
reverting) dynamics corresponding to $a>0$ and $b=d=e=0$.

In the Econophysics literature several studies have been devoted to asses the
statistical properties of the volatility (see for instance Chapter 7
in~\cite{Bouchaud_book:2000} and~\cite{Micciche:2002}), especially its
distribution, and it has been recognized that the instantaneous volatiliy,
measured by suitable proxies, distributes in good agreement with a Log-Normal
or an Inverse Gamma law, the best fit being obtained with the
latter~\cite{Bouchaud_book:2000} which is able to better capture the heavy tail
of the empirical distribution. None of the previously cited models feature an
Inverse Gamma probability density function (PDF) for $\sigma_t$, even though
this distribution has been considered previously in different contexts. For
instance, the Inverse Gamma was introduced in the context of an ARCH-like
evolution of the variance in~\cite{JEconometrics.45.7}, and in the statistical
modeling of financial data the marginalization of Normally distributed returns
conditionally on Inverse Gamma variance was widely exploited since it generates
generalized Student-$t$ distributions (see~\cite{JBusiness.45.49,PhysRevE.80.65102}).
However, as clarified by the empirical analysis performed
in~\cite{Micciche:2002}, where intra-day returns are used to estimate a proxy
for the daily volatility, an Inverse Gamma PDF for $\sigma^2_t$ leads to an
overweighting of the tail region.

Here we afford the statistical characterization of the simplest linear SVM able
to account for this stylized fact about the volatility. The
process~\eqref{eq:Y_MNDP} has been extensively studied and characterized
in~\cite{PhysRevE.81.032102} where exact solutions for the moments of the
associated PDF have been obtained allowing to study its relaxation modes toward
a stationary distribution, if any. In particular, when $a<0$ and $d=e=0$, with
$c>0$, the process~\eqref{eq:Y_MNDP} has indeed an Inverse Gamma stationary
distribution, whose support is $[0,+\infty)$ as long as $b>0$. Thereby we
consider the following SVM
\begin{equation}
  \begin{aligned}
    dX_t &= \sqrt{c}\, Y_t \,dW_{1,t}\,, \quad X_{0} = 0\\
    dY_t &= \left( a Y_t + b \right)\,dt + \sqrt{c} \,Y_t \,dW_{2,t}\,,
    \quad Y_{t_0}=y_{t_0}\,,
  \end{aligned}
  \label{eq:IGmodel}
\end{equation}
where $t_0 \leq 0$, $y_{t_0}$ may be a fixed constant or randomly sampled, and
the constant factor in the expression of the instantaneous volatility
$\sigma_t=\sqrt{c}\,Y_t$ has been added for later convenience. As explained
in~\cite{PhysRevE.81.032102} the stationary
PDF of $\sigma_t$ is
\begin{equation}
  \Pi_{st}(\sigma) = \frac{\lambda^{\nu}}{\Gamma(\nu)}
  \,\frac{\exp{\left(-\lambda/\sigma\right)}}{\sigma^{\nu+1}} \,,
  \label{eq:sigmaDist}
\end{equation}
where the shape parameter $\nu$ and the scale parameter $\lambda$ are
given by
\begin{equation}
	\nu=1-\frac{2 a}{c} \quad\mathrm{and}\quad \lambda = \frac{2 b}{\sqrt{c}}\,.
  \label{eq:shapeScale}
\end{equation}

\section{Emergence of fat tails} \label{sec:fatTails}

A major point to be discussed before presenting a detailed derivation of our
results is the different role played by the initial time conditions for the $X$
and $Y$ processes.
Since $X_t$ represents the detrended logarithmic increment of the price over
the time lag $t$, it can be directly measured from real time series, and in
a natural way we can assume as starting point for this process the spot time
$t=0$.
On the other hand, the secondary process can not be observed directly but some
of its statistical properties have been measured by means of suitable proxies.
In particular, for intra-day frequencies there is no clear evidence of mean
reversion, that is the high frequency volatility is very close to its
asymptotic value~\cite{QuantFinance.4.176,QuantFinance.6.115}. In order to
capture this evidence, we assume that the process $Y$, driving the returns from
$0$ to $t$, started in the past at $t_0<0$ and we will perform the limit $t_0\to -\infty$ at the end.
The assumption of stationarity for the $\sigma_t$ process in~\eqref{eq:dXeq}
allows also to consider the returns $dX_t$ as identically distributed and
uncorrelated, even though not independent variables, by virtue of the
\emph{i.i.d.} property of the Wiener increments.

The structure of the model~\eqref{eq:IGmodel} allows to compute the
moments of the PDF of $X_t$ at all times $t$ recursively. Application of the
It\^o Lemma to the function $X_t^n$ readily provides
\begin{equation*}
  \ang{X_t^n} =
  \frac{1}{2} n (n-1) c \int_{0}^{t} \ang{X_s^{n-2}\, Y_s^{2}}\,ds \,,
\end{equation*}
and the same Lemma proves that the correlation functions between $X$
and $Y$ satisfy the following differential equation
\begin{multline}
  \frac{d}{dt} \ang{X_t^p Y_t^q} = F_q \,\ang{X_t^p Y_t^q}
  + A_q \,\ang{X_t^p Y_t^{q-1}}\\
  +c \,\rho \,p \,q \,\ang{X_t^{p-1} Y_t^{q+1}}
  + \frac{1}{2} p (p-1)c \,\ang{X_t^{p-2} Y_t^{q+2}} \,,
  \label{eq:corrDiffEq}
\end{multline}
where we defined $F_k=k a + k(k-1) c /2$, $A_k=k b$ for every $k\in\mathbb{N}$,
and $p,q\in \mathbb{N}$. The previous equation is a linear ordinary
differential equation (ODE) for every $p$ and $q$, which can be solved recursively
starting from the lowest order of $p$ and $q$~\footnote{It is worth mentioning that a similar
equation holds for the more general dynamics~(3), after defining the volatility
as $\sigma_t=\sqrt{c \,Y_t^2 + d \,Y_t + e}$.}, and whose solution involves
integration of the moments $\ang{Y_t^n}\doteq \mu_n(t;t_0)$ of the $Y$ process.
For every $n$ and every time $t$ the latter can be expressed as a linear
superposition of exponential functions
\begin{equation}
  \mu_n(t;t_0) = \sum_{j=0}^n K^{(n)}_j \,\exp{\left[F_j (t-t_0)\right]} \,.
  \label{eq:Ymoms}
\end{equation}
The explicit expressions of the coefficients in the above expansion can be
computed as explained in~\cite{PhysRevE.81.032102}, and it turns out that
$K^{(n)}_j$ involves the values $\mu_{k}(t_0;t_0)$ for $k=1,\dots, j$, while $K^{(n)}_0$ does not. 
This implies that whenever the constants $F_j$ are all negative, the only term
surviving in the limit $t_0\to -\infty$ is $K^{(n)}_0$ and the process looses
every information about the distribution of $y_{t_0}$. It is worth noticing that,
even though the moments $\mu_n(t;t_0)$ are homogeneous functions of time, 
when $t_0$ is finite this is not true for the solution of
Eq.~\eqref{eq:corrDiffEq} which is obtained by integration from $0$ to $t$,
with boundary condition $\ang{X_0^p \,Y_0^q}=0$ for every $p>0$~\footnote{
From now on we will drop the dependence on $t_0$ from the moments
$\mu_n$.}.

From the analysis of Eq.~\eqref{eq:corrDiffEq} it can be verified
that the moments of $X$ can be expressed always as a superposition of exponential
functions of the starting time of the volatility as follows
\begin{equation}
  \ang{X^n_t} = \sum_{j=0}^{n} H^{(n)}_j(t) \,\exp{\left(-F_j t_0\right)} \,.
  \label{eq:XmomsExpand}
\end{equation}
The coefficients $H^{(n)}_j$ depend on the time lag $t$ and, more
precisely, by virtue of the linearity of the ODEs~\eqref{eq:corrDiffEq}, they
correspond to a combination of exponential terms weighted by polinomial
functions of $t$. In Appendix~\ref{app:X3} we report the explicit expressions
of the coefficients $H^{(n)}_j(t)$ for the cases $n=2$ and $n=3$, from which it
can be readily verified that the skewness of the PDF converges to zero
asymptotically for $t\to +\infty$. A messy calculation would show that an
analogous behavior holds for kurtosis. Thus the scaling of the lowest order
moments is in full agreement with the one of the empirical distributions over
long time horizons~\cite{Mantegna_book,Bouchaud_book:2000}. 
When $t$ is finite the coefficients $H^{(n)}_j$ are finite
quantities themselves, and all the relevant information about the behavior of
$\ang{X^n_t}$ in the stationary limit of $Y$ is retained by the
$t_0$-exponentials in Eq.~\eqref{eq:XmomsExpand}. Two cases are possible here:
if all the $F_j$ are negative ($j\neq 0$), $\ang{X^n_t}$ is finite in the
stationary limit $t_0\to-\infty$, otherwise it diverges~\footnote{Since $F_j
\neq F_k$ for every $j,k>1$ with $j\neq k$, no cancellation of the divergent
terms can take place in the limit $t_0\to -\infty$.} indicating the emergence
of fat tails in the PDF of $X_t$. The latter case applies when $n > \nu = 1-2
\,a/c$, as can be checked directly from the definition of $F_n$. Since
$F_{n+1}>F_n$, when $F_n>0$, the divergence of $\ang{X^n_t}$ implies the
divergence of all the higher order moments~\footnote{The case $\rho=0$
represents an exception since, due to symmetry arguments, all the odd moments
vanish identically.}.
The same condition is responsible for the divergence of the moments $\mu_n(t)$
of the volatility for $n>\nu$ (see Eq.~\eqref{eq:Ymoms}) in agreement with the
fact that the stationary distribution of the volatility~\eqref{eq:sigmaDist}
is an Inverse Gamma distribution with tail index $\nu$.
Here we see at work a mechanism in which the power law tail of the stationary
distribution of the volatility induces fat tails in
the return distribution for every time lag $t$, and its scaling for large $\abs{x}$ is compatible with
a power law assumption
\begin{equation*}
	p(x) \underset{x\to \pm \infty}{\thicksim} \frac{1}{\abs{x}^{1+\beta}}\,.
\end{equation*}
This is in agreement with empirical studies about the distribution of returns
over daily or intra-day time
scales~\cite{PleGop1999,Bouchaud_book:2000,gell2004nonextensive,PhysA.324.89,Mantegna_book},
and from the previous considerations we are able to constraint the tail index
in the following range
\begin{equation}
  n^* < \beta \le n^*+1 \,,
  \label{eq:pdfTailIndex}
\end{equation}
where $n^*>0$ is the largest integer satisfying $n^*<\nu$.
\begin{figure}[tb]
  \begin{center}
    \includegraphics[scale=0.7]{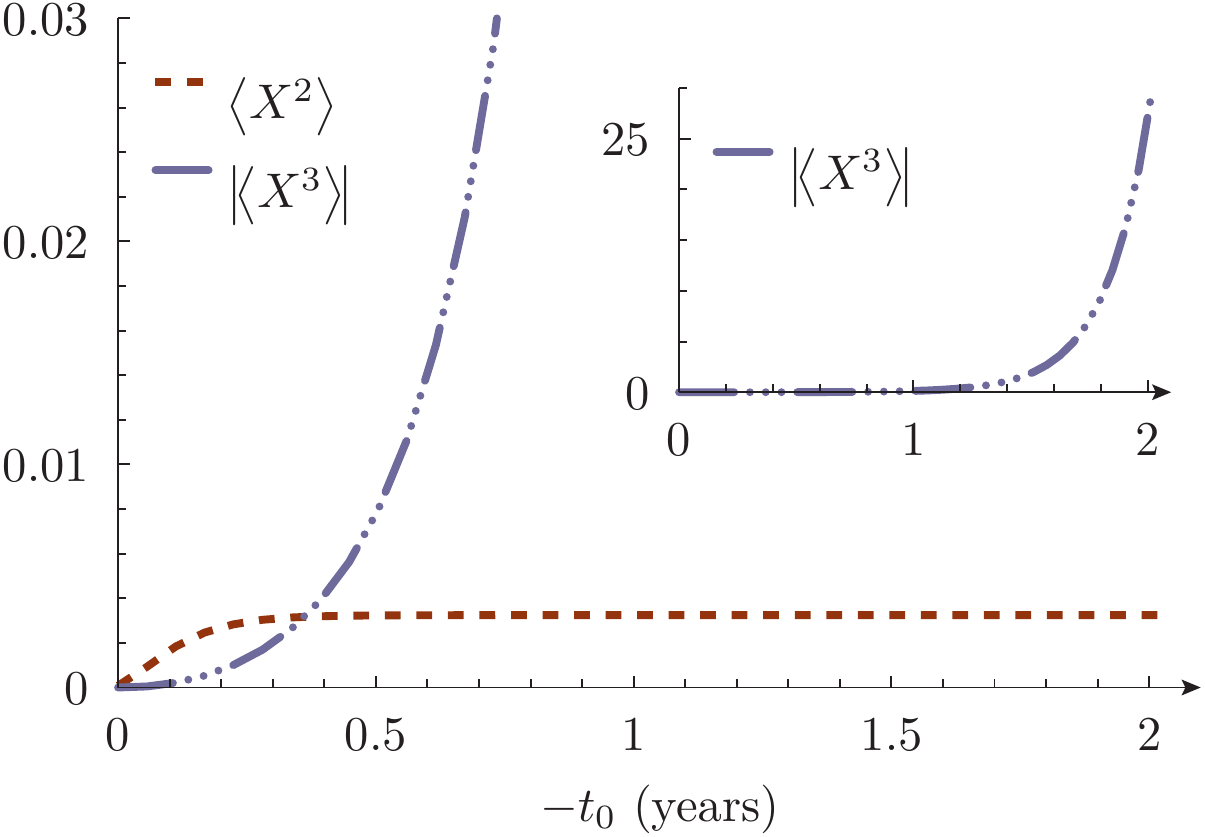}
  \end{center}
  \caption{(Color online) Scaling as a function of $t_0$ of the second and third moment of $X$
  at $t=1~\mathrm{day}$, for $a=-16.06~\mathrm{yr}$, $b=0.86~\mathrm{yr}$,
  $c=17.84~\mathrm{yr}$ and $\rho=-0.51$, $\abs{a}/c=0.6$. Yearly units
  (1 yr = 250 trading days).}
  \label{fig:t0scaling}
\end{figure}
As an example, in Fig.~\ref{fig:t0scaling} it is shown the scaling of
$\ang{X^2_t}$ and of the absolute value of $\ang{X^3_t}$ as a function of the
starting time of the volatility, for $t=1~\mathrm{day}$ and for a choice of the
parameters corresponding to $\abs{a}/c=0.6$. For this value of the ratio the
tail index of the return distribution is $2<\beta\le 3$ and consequently the
third moment of the stationary distribution of the volatility diverges as $t_0$
becomes more and more negative, while $\ang{X^2_t}$ approaches its finite
stationary value.

\section{Leverage effect}\label{sec:lev}

For the linear model~\eqref{eq:IGmodel} the leverage,
measuring the correlation between returns and volatility, can be
computed exactly. Since the squared increment $d X^2$ provides an estimation of the
instantaneous volatility, it can be defined through the following function
\begin{equation}
  \mathcal{L}(\tau;t) = \frac{\ang{dX_t \,dX_{t+\tau}^2}}{\ang{dX_t^2}^2} \,.
  \label{eq:levDef}
\end{equation}
Empirically, for arbitrary $t$, $\mathcal{L}(\tau; t)$ is found to be exponentially decaying for
positive $\tau$ and approximately zero otherwise, meaning that a
correlation exists between past returns and the volatility in the future
and not \emph{vice versa}. Empirical analysis shows that it is a short range
correlation; more precisely, the decay time of $\Lc(\tau; t)$ is found to be
of approximately $69$~days for U.S. stocks and even smaller, about 10~days,
for indexes~\cite{Bouchaud_book:2000}.

The numerator~\eqref{eq:levDef} can be rewritten as
\begin{equation*}
  \ang{dX_t \,dX_{t+\tau}^2} = c^{3/2} \ang{\zeta_{1,t} \,Y_t \,Y_{t+\tau}^2}\,dt^2 \,,
\end{equation*}
expressing the Wiener increment as $\zeta_t \,dt$, where
$\zeta_t$ is a Gaussian noise with zero mean and $1/dt$ variance.
Novikov theorem~\cite{SovPhys.20.1290,PhysRevE.67.037102} allows to
compute the expectation involving $\zeta_{1,t}$, giving us
\begin{multline*}
  \ang{dX_t \,dX_{t+\tau}^2} = \\
  2 \rho\,c^2 H(\tau)\exp{\left(a \tau\right)}
  \ang{Y_t^2 \,Y_{t+\tau} \, \exp{\left[\sqrt{c} \Delta_t W_2(\tau)\right]}}\,,
\end{multline*}
where we defined $ \Delta_t W(\tau) \doteq \int_{t}^{t+\tau}\,dW_s$,
we took into account the correlation structure~\eqref{eq:Wcorrel} and
we used the following expression of the functional derivative of $Y$
\begin{multline*}
  \frac{\delta Y_{t+\tau}}{\delta \zeta_{1,t}} = \rho \,\frac{\delta Y_{t+\tau}}{\delta \zeta_{2,t}} = \\
  \rho \sqrt{c} H(\tau) \exp{\left(a \tau\right)} Y_t \exp{\left[\sqrt{c} \Delta W_{2,t}(\tau)\right]} \,,
\end{multline*}
with the Heaviside step function $H(\tau)$ defined as zero if $\tau\leq 0$ and one otherwise. 
The expectation $f(\tau;t,Y)\doteq \ang{Y_t^2 \,Y_{t+\tau} \exp{\left[\sqrt{c}\,\Delta_t W_2(\tau)\right]}}$ 
satisfies an integral Volterra equation of the
second kind, whose derivation is detailed in Appendix~\ref{app:leverage}, and
the final expression of the leverage correlation reads
\begin{multline}
  \mathcal{L}(\tau;t) = \frac{2 \,\rho\, H(\tau)}{\mu_2(t)^2}
  \left\{ \left[ \mu_3(t) + \frac{b}{a+c} \,\mu_2(t) \right] \times \right.\\
  \left.\exp{\left[\left( 2 a + \frac{3}{2} c \right)\tau\right]} -\frac{b}{a+c} 
  \,\mu_2(t) \exp{\left[\left( a+\frac{c}{2} \tau \right)\right]}\right\} \,,
  \label{eq:lev3}
\end{multline}
which inherits the explicit dependence on $t$ from the moments of $Y$. In
order to compare the previous expression with real data, following the
discussion at the beginning of Section~\ref{sec:fatTails}, we take the limit
$t_0\to -\infty$, so that we can replace $\mu_2(t)$ and $\mu_3(t)$ with
their asymptotic values, whose general expression, valid for $n<\nu$, is
\begin{equation}
  \mu_{n,st} = K^{(n)}_0 = \prod_{k=1}^n (-1)^k\,\frac{A_k}{F_k} \,.
  \label{eq:YstatMoms}
\end{equation}
Substitution in Eq.~\eqref{eq:lev3} reveals that the first term vanishes,
and the leverage correlation reduces to
\begin{equation}
  \mathcal{L}(\tau) = -\rho \,H(\tau) \,\frac{a (2 a + c)}{b (a+c)}
  \exp{\left(-\frac{\tau}{\tau^\mathcal{L}}\right)} \,,
  \label{eq:levFinal}
\end{equation}
where the leverage decay time reads
\begin{equation*}
  \tau^{\mathcal{L}}= \frac{2}{2 \abs{a}-c} \,.
\end{equation*}
So, the model correctly forecasts the exponential decay of $\mathcal{L}(\tau)$
and its vanishing for negative correlation times.

\section{Volatility autocorrelation}\label{sec:autoc}

The volatility autocorrelation provides an estimate of how much the
volatility at time $t+\tau$ depends on the value it had at time $t$ and it
is usually defined as
\begin{equation}
  \Ac(\tau;t) = \frac{\ang{dX_t^2 \,dX_{t+\tau}^2}-\ang{dX_t^2}
  \ang{dX_{t+\tau}^2}}{\sqrt{\var[dX_t^2]\, \var[dX_{t+\tau}^2]}}\,.
  \label{eq:AcorrDef}
\end{equation}
It is a well known stylized
fact~\cite{MP_Qfin:2006,RePEc:taf:apmtfi:v:11:y:2004:i:1:p:27-50,Perello2004134}
that $\Ac$ decays with multiple time scales and in particular, it shows
a long range memory effect, vanishing over a time scale of the order
of few years for stock indexes.

For the model under investigation, the volatility autocorrelation can
be computed exactly too.  Recalling again the Novikov theorem and the fact that
$\delta \,dW_{1,t}/\delta \zeta_{1,t}=1$, the correlation entering the
numerator of~\eqref{eq:AcorrDef} becomes
\begin{multline*}
  \ang{dX_t^2 \,dX_{t+\tau}^2} = c^2 \ang{Y_t^2 \,Y_{t+\tau}^2} \,dt^2 \\ 
  +2\,\rho c^{5/2} H(\tau) \ang{ Y_t^2 \,Y_{t+\tau} \exp{\left[\sqrt{c} \,\Delta_tW_2(\tau)\right]} 
  \,dW_{1,t}} \,dt^2 \,,
\end{multline*}
but, due to the presence of $dW_{1,t}$, the second term results to be of
order $\mathcal{O}(dt^3)$ and therefore it can be discarded. The exact
expression of the autocorrelation function $\ang{Y_t^2 \,Y_{t+\tau}^2}$ can
be obtained as explained in Appendix~\ref{app:autocorr}, leaving us with
\begin{widetext}
  \begin{equation*}
    \Ac(\tau;t) \!=\! \frac{\exp{\left(a\tau\right)}}{3\mu_4(t)-\mu_2(t)^2}\!\left\{ 
    \frac{2b}{a+c} \left[ \mu_1(t) \mu_2(t) \!-\! \mu_3(t)\right]+\exp{\left[(a+c)\tau\right]}
    \left[ \mu_4(t) \!+\! \frac{2b}{a+c} \mu_3(t) \!-\! \mu_2(t)\left(
    \mu_2(t) \!+\! \frac{2b}{a+c} \mu_1(t) \right) \right] \right\}\,, 
  \end{equation*}
\end{widetext}
where the denominator of Eq.~\eqref{eq:AcorrDef} has been approximated with
$\var[dX_t^2]=c^2\left[ 3\mu_4(t)-\mu_2(t)^2 \right]\,dt^2$ in view of the
stationary limit for $Y$. After replacing the moments $\mu_n(t)$ with their
asymptotic expressions~\eqref{eq:YstatMoms} we end with
\begin{equation}
  \Ac(\tau) = \frac{1}{D} \left[ N_1 e^{-\tau/\tau^{\mathcal{A}}_1}+N_2
  e^{-\tau/\tau^{\mathcal{A}}_2} \right] \,,
  \label{eq:AcorrFinal}
\end{equation}
where the coefficients read
\begin{align*}
  D&= \frac{\left( 4 a^2-2 a c - 3 c^2 \right)\left( a + c \right)}{c^2}\\
  N_1&=-\frac{\left( 2a + 3c \right)\left( 2a + c \right)}{c}\\
  N_2&=a \,,
\end{align*}
and we also defined the two volatility autocorrelation time scales as
\begin{equation*}
	\tau^{\mathcal{A}}_1=\frac{1}{|a|}\quad\mathrm{and}\quad
	\tau^{\mathcal{A}}_2=\frac{1}{2|a|-c} \,.
\end{equation*}

At this point it is crucial to notice that in deriving
Eq.~\eqref{eq:levFinal} and Eq.~\eqref{eq:AcorrFinal} we assumed implicitly that
the moments of $Y_t$ up to the order $n=4$ do converge asymptotically.
Recalling the expression of the shape parameter $\nu$ in~\eqref{eq:shapeScale},
we see this assumption imposes
\begin{equation}
  \frac{\abs{a}}{c} > \frac{3}{2} \,,
  \label{eq:ratioBound}
\end{equation}
which has to be interpreted as a consistency relation for the model.  This
constraint imposes the following strict ordering between the time scales of
the model
\begin{equation}
  \tau^{\mathcal{A}}_2 < \tau^{\mathcal{A}}_1 < \tau^{\mathcal{L}}\,,\quad
  \text{with}\quad \tau^{\mathcal{A}}_1> \frac{2}{3}\tau^{\mathcal{L}}~,
  \label{eq:ordering}
\end{equation}
where the second inequality for $\tau_1^{\Ac}$ follows from the convergence of
third moment of $Y_t$ which requires $\abs{a}/c > 1$.

The expression obtained for $\Ac$ fails to capture the persistence
of this correlation identified in several analysis reviewed in~\cite{Qfin.1.223}.
The lacking of power law scaling would not be, in principle, a serious drawback
as far as one of the two time scales involved in~\eqref{eq:AcorrFinal} was
sufficiently long. However, the ordering~\eqref{eq:ordering}, which is peculiar of the
considered model, makes these scales too close each other and the volatility
autocorrelation to decay as fast as $\Lc$, an undesired feature
shared with other models, such as the Stein-Stein one.
The persistence of $\Ac$ can be accounted for introducing a non linear
volatility, as it is for the exponential Ornstein-Uhlenbeck
model~\cite{MP_Qfin:2006}, or coupling a third stochastic equation driving the
dynamics of the long run value of $Y_t$ as
in~\cite{RePEc:taf:apmtfi:v:11:y:2004:i:1:p:27-50}. A further possibility to
induce a non exponential time decay would be to consider a non linear drift term
for the dynamics of $Y_t$, even though the analytical tractability of the
present model will not be preserved.

\section{Estimation of parameters}\label{sec:estimation}
\begin{figure}[b]
  \begin{center}
    \includegraphics[scale=0.7]{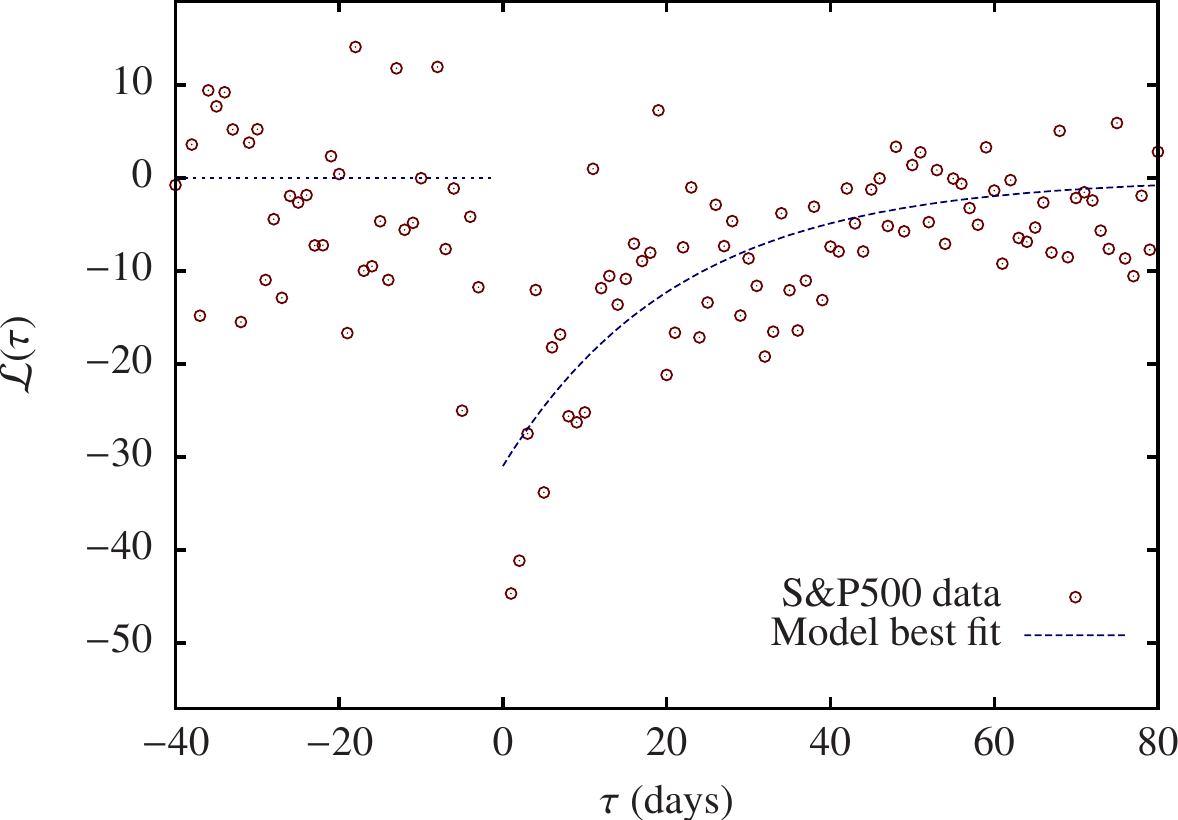}
  \end{center}
  \caption{(Color online) Best fit of the empirical leverage correlation with the
  model prediction~\eqref{eq:levFinal} as a function of the two
  parameters $\tau^{\Lc}$ and $\Lc(0^+)$.}
  \label{fig:LevFit}
\end{figure}

Now we provide a systematic methodology for estimating the model parameters, 
which are the constants $a$, $b$, $c$ entering the dynamics
of $Y_t$, plus the correlation coefficient $\rho$. We perform the
estimation over the Standard \& Poor 500 (S\&P500) index daily returns from 1970 to
2010, approximating $dX_t$ with $\Delta X_t = X_{t+\Delta t}-X_t$
\begin{equation*}
  dX_t \approx \Delta X_t = \ln\left(\frac{S_{t+\Delta t}}{S_t} \right)
  -\ang{\ln\left(\frac{S_{t+\Delta t}}{S_t} \right)} \,,
\end{equation*}
where $\Delta t=1/250~\mathrm{yr}$ (one trading day). Taking into account
that $dW_{1,t}$ is independent of $\sigma_t$ and that $\abs{\Delta W_1}$ is
distributed accordingly to a Folded Normal law, the following relations
hold for the model~\eqref{eq:IGmodel}
\begin{align*}
  A &\doteq \frac{\ang{\abs{\Delta X}}}{\ang{\abs{\Delta W_1}}} =
  \sqrt{\frac{\pi}{2 \Delta t}} \ang{\abs{\Delta X}} =-\sqrt{c}
  \,\frac{b}{a}\\
  B & \doteq \frac{\ang{\Delta X^2}}{\ang{\Delta W_1^2}} =
  \frac{\ang{\Delta X^2}}{\Delta t}  =c \,\frac{2 b^2}{(2 a + c) a} \\
  C &\doteq \frac{\ang{\abs{\Delta X}^3}}{\ang{\abs{\Delta W_1}^3}} =
  \sqrt{\frac{\pi}{(2 \Delta t)^3}} \,\ang{\abs{\Delta X}^3}\\
  &= -\frac{2 b^3 \,c^{3/2}}{(a+c)\,(2a+c)\,a}  \,.
\end{align*}
The constants $A$ and $B$ can be measured directly from the data, providing
us an estimation of the ratio $a/c$ through the relation
\begin{equation*}
  D \doteq \frac{B}{2 \left( A^2-B \right)} = \frac{a}{c} \,.
\end{equation*}
The value of these quantities extracted from the series of the daily
returns of the S\&P500 index are reported in Table~\ref{tab:spotValues}.
It is crucial to observe that the value obtained for the ratio $\abs{a}/c$
is compatible with the constraint~\eqref{eq:ratioBound}, supporting the
consistency of our model and the convergence of the volatility
autocorrelation. Moreover, the same ratio provides an estimate of $\nu=4.579$
(see Eq.~  \eqref{eq:shapeScale}) implying for the order of the highest
converging moment a value $n^*=4$; consequently,
relation~\eqref{eq:pdfTailIndex} indicates the following range for the tail
index of $p(x)$
\begin{equation*}
  4 < \beta \le 5 \,.
\end{equation*}
The leverage correlation~\eqref{eq:levFinal} provides a
way to obtain the two further relations needed to fix the four free
parameters of the models. Indeed, a two parameters fit of the function
$\Lc(\tau)$ gives estimates for the time scale $\tau^{\Lc}$ and for the
limit $\tau\to 0^+$
\begin{equation*}
  \Lc(0^+) \doteq -\rho \,\frac{a (2a + c)}{b (a+c)} \,,
\end{equation*}
with the results reported in Table~\ref{tab:fitLev} and
Fig.~\ref{fig:LevFit}.
% TABLE 1
\begin{table}
  \centering
  \begin{tabular}{cccl}
    \toprule
    Estimators	&&	\multicolumn{2}{c}{S\&P500 daily returns}\\
    \colrule
    A & &0.1457  & $\mathrm{yr}^{-1/2}$\\
    B & &0.0295  & $\mathrm{yr}^{-1}$\\
    C & &0.0107  & $\mathrm{yr}^{-3/2}$\\
    $\abs{a}/c$ & &1.7895 &\\
    \botrule
    \hline
  \end{tabular}
  \caption{Estimates from return sample averages. We compute the value of the
  estimators $A$, $B$, $C$ and $D$ for the daily log-returns of the S\&P500
  index during the period 1970-2010, exploiting the means of $\abs{\Delta X}$,
  $\Delta X^2$ and $\abs{\Delta X}^3$.}
  \label{tab:spotValues}
\end{table}
% FIGURE 1
In particular, the value obtained for the leverage time scale,
$\tau^{\Lc}\approx 21~\mathrm{days}$, and for its amplitude
$\mathcal{L}(0^+)$ are consistent with those quoted in past analysis of
different stock indexes such as the Dow Jones Industrial
Average~\cite{MP:2002,RePEc:taf:apmtfi:v:11:y:2004:i:1:p:27-50}, and
confirm the short range nature of this effect.
% TABLE 2
\begin{table}
  \centering
  \begin{tabular}{ccrl}
    \toprule
    Estimators	&&	\multicolumn{2}{c}{S\&P500 daily returns}\\
    \colrule
    $\tau^{\Lc}$ && 0.0864 & yr \\
    $\Lc(0^+)$   && -30.9515 &  \\
    \botrule
  \end{tabular}
  \caption{Estimation of the leverage time scale and its limit for $\tau\to
  0$, obtained from the fit of the empirical leverage
  correlation~\eqref{eq:levDef} for the daily log-returns of the S\&P500
  index, with the model predicted expression~\eqref{eq:levFinal}.}
  \label{tab:fitLev}
\end{table}

\begin{figure}
  \begin{center}
    \includegraphics[scale=0.7]{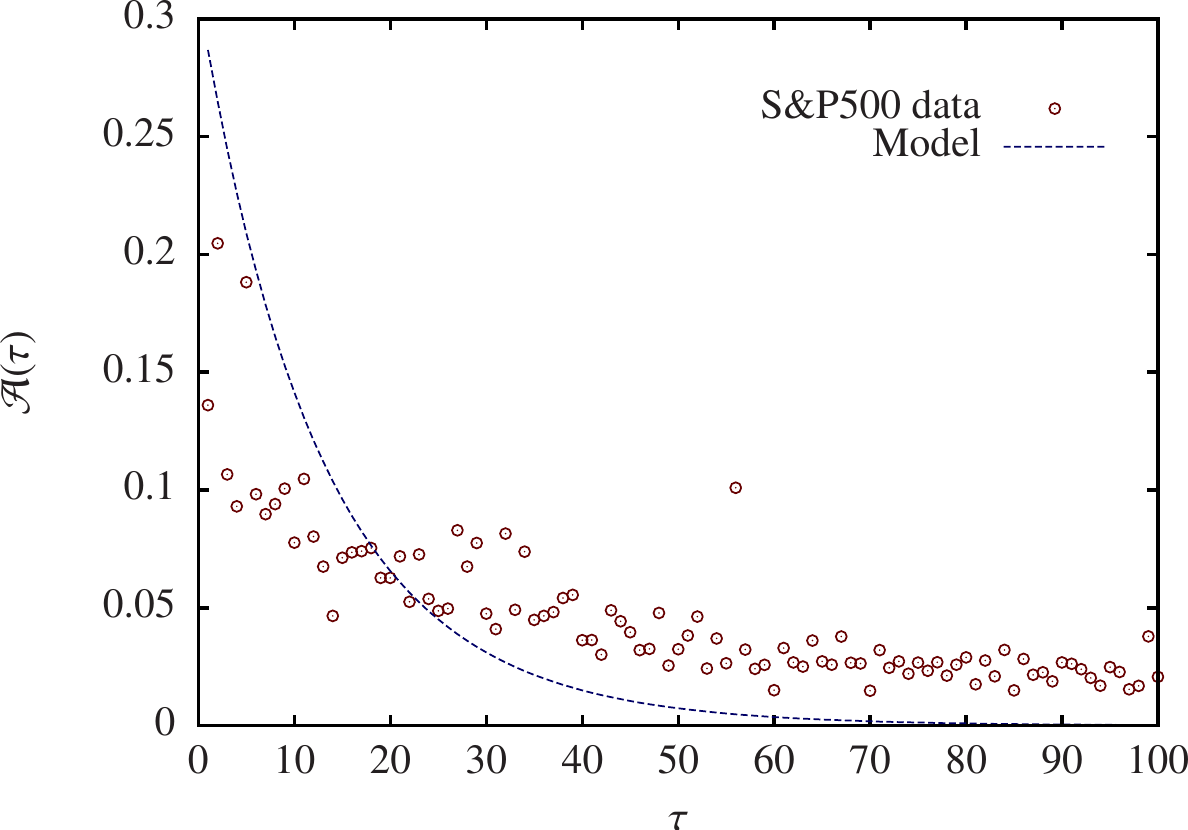}
  \end{center}
  \caption{(Color online) Theoretical prediction for the volatility autocorrelation
  function of the daily returns of the S\&P500 index 1970-2010,
  Eq.~\eqref{eq:AcorrFinal}.}
  \label{fig:AcorrTh}
\end{figure}

At this point all the parameters can be recovered through the following
relations
\begin{align}
  \label{eq:paramsFinala}
  c &= -\left[ \tau^{\Lc}\left( D+\frac{1}{2} \right) \right]^{-1} \\
  a &= c\, D \\
  b &= -\frac{a+c}{\sqrt{c}} \,\frac{C}{B} \\
  \label{eq:paramsFinalrho}
  \rho &= - \frac{ b \,(a+c)}{a \,(2 a + c)} \,\Lc(0^+) \,.
\end{align}
% TABLE 3
\begin{table}
  \centering
  \begin{tabular}{ccrl}
    \toprule
    Parameter & & \multicolumn{2}{c}{Estimate from S\&P500} \\
    \colrule
    $a$     & & $-16.0608$ &   $\mathrm{yr}^{-1}$ \\
    $b$     & & $0.8627$   &   $\mathrm{yr}^{-1}$ \\
    $c$     & & $8.9749$   &   $\mathrm{yr}^{-1}$ \\
    $\rho$  & & $-0.5089$  &   			  \\
    \botrule
  \end{tabular}
  \caption{Model parameters estimated from the daily log-returns of the
  S\&P500 index during 1970-2010 through the
  relations~\eqref{eq:paramsFinala}-\eqref{eq:paramsFinalrho}.}
  \label{tab:parsFinal}
\end{table}
The final results, reported in Table~\ref{tab:parsFinal}, show a negative
correlation coefficient, in agreement with the known leftward asymmetry of
daily return distributions. Moreover, our calibration provides for the
relaxation time of the volatility process a finite value $\tau^{\sigma} \doteq -1/a \approx
15~\mathrm{days}$, implying that, from a practical point of view, the limit $t_0\to -\infty$
is equivalent to $t_0 \ll -\tau^{\sigma}$.
The fitted values of $\tau^{\Lc}$ and $\Lc(0^+)$ provides a good
description of real data, as shown in Fig.~\ref{fig:LevFit}; on the other
hand, Fig.~\ref{fig:AcorrTh} shows that the theoretical
volatility autocorrelation for the estimated values of the parameters,
Eq.~\eqref{eq:AcorrFinal}, does not capture the long range persistence
of the empirical volatility, as expected from the constraints~\eqref{eq:ordering},
while it describes correctly the exponential decay for small values of $\tau$.

\begin{figure}[tb]
  \begin{center}
    \includegraphics[scale=0.7]{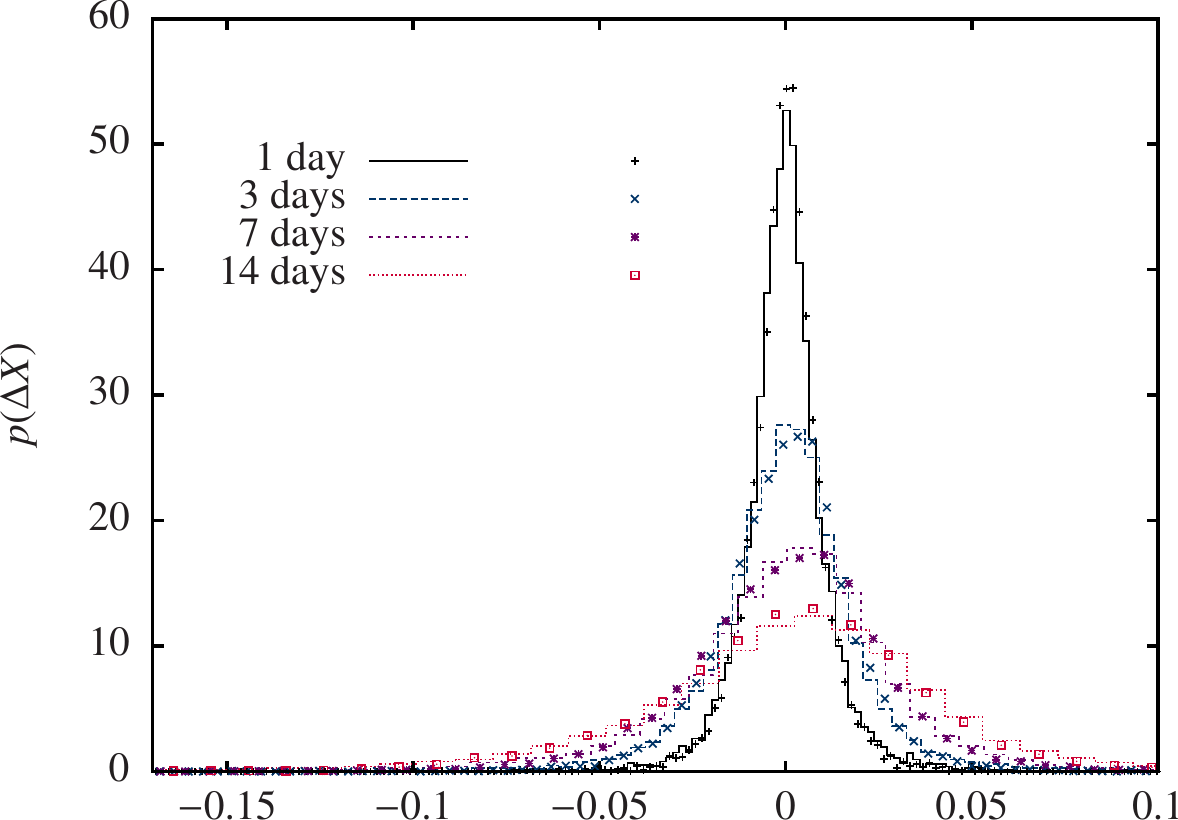}
  \end{center}
  \caption{(Color online) Linear plot showing the comparison between the return PDFs
  predicted by the model (lines) and the data for the S\&P500 index,
  for different time scales.}
  \label{fig:linPDF}
\end{figure}

\begin{figure}[tb]
  \begin{center}
    \includegraphics[scale=0.7]{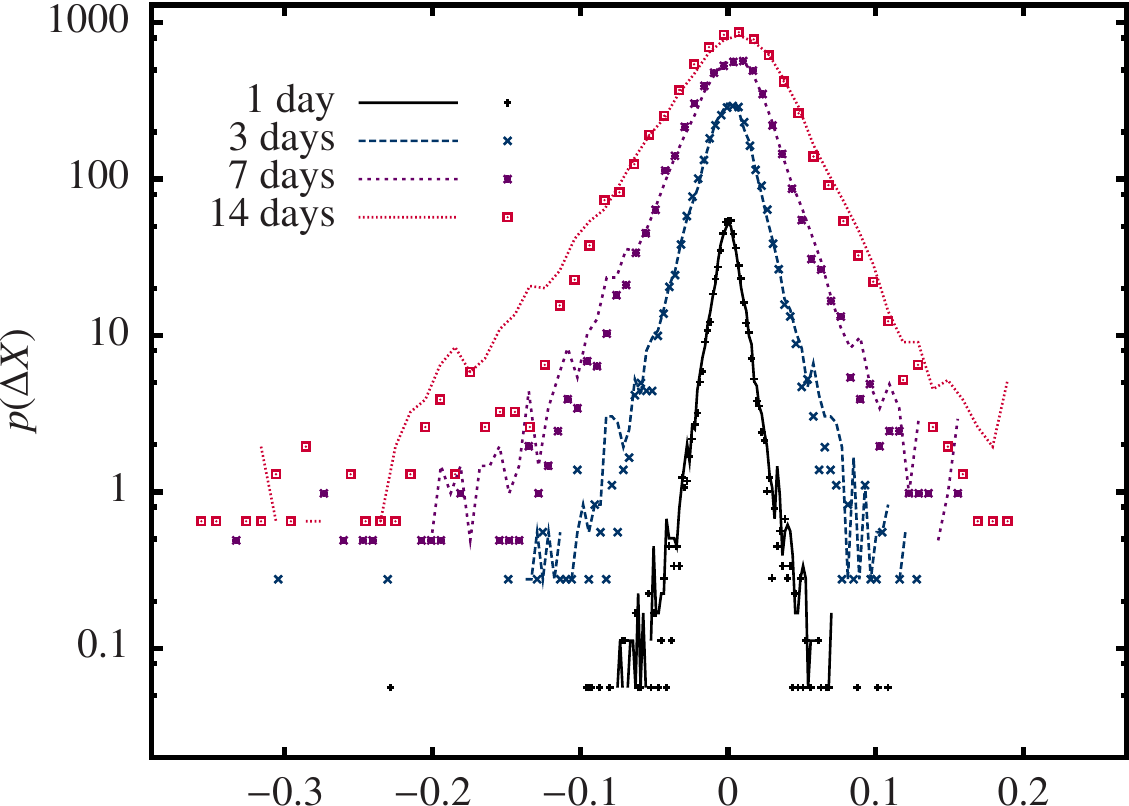}
  \end{center}
  \caption{(Color online) In log-linear scale, return probabilities for the model (lines)
  \emph{vs} S\&P500 returns (points). Curves have been shifted for sake
  of readability.}
  \label{fig:linLogPDF}
\end{figure}

Finally, it is important to compare the return PDF predicted by the
model with the data sample from which the model parameters were estimated.
Since we model the return dynamics for increasing $t$, it is even more important to asses to which
extent the diffusion process~\eqref{eq:IGmodel} is able to capture the
scaling properties of the empirical distribution over different time
horizons. At this aim, with the parameters fixed from the daily S\&P500
series, we reconstruct the theoretical PDFs simulating the process at
different time scales, $t = 1,3,7,14~\mathrm{days}$, and we compare 
them with the corresponding empirical distributions obtained aggregating
the daily returns. This comparison is shown in
Fig.~\ref{fig:linPDF} and Fig.~\ref{fig:linLogPDF}. The daily distribution is very
well reproduced by the theoretical PDF, which is able to fully capture the
leptokurtic nature of the daily data. The plots also confirm that the
diffusive dynamics~\eqref{eq:IGmodel}, once the parameters have been fixed at the
daily scale, follows closely the evolution of the empirical curves for
larger $t$.
In particular, it captures the progressive convergence in the central region
to a distribution with vanishing skewness and kurtosis.

\section{Conclusions}\label{sec:concl}

In this work, we have introduced a class of SVMs where the volatility is driven
by the general process with multiplicative noise analyzed in detail
in~\cite{PhysRevE.81.032102}. More specifically, we focused on the set of
parameters resulting in an Inverse Gamma stationary distribution for the
$\sigma_t$ process.  We provided an analytical characterization of the moments
of the return distribution, revealing the role played by the power law behavior
of the Inverse Gamma in the emergence of fat tails. Nevertheless, even though
the highest order moments of $X$ diverge for every time lag, the analytical
expressions we obtained reveal the vanishing of both the skewness and the
kurtosis, in agreement with the Normality of returns for long horizons. As far
as the estimation procedure is concerned, it is worth noticing that we do not
exploit directly the statistical properties of the instantaneous volatility
which is an hidden process, but on the contrary we infer the Inverse Gamma
parameters from well established robust stylized facts holding at the daily
scale. Indeed our model correctly predicts zero autocorrelation for the
returns, and the short range exponential decay of the leverage.
The persistence of the volatility autocorrelation over yearly horizons is not
captured, and in this perspective we would like to explore the possibility of
coupling a third SDE in the same spirit
of~\cite{RePEc:taf:apmtfi:v:11:y:2004:i:1:p:27-50}. Moreover, we expect that
relaxing the time homogeneity of the processes, as done
in~\cite{PhysRevE.81.032102}, we may induce time scalings more general than the
exponential one. We also expect the proposed dynamics to be a good candidate to
describe the price and volatility dynamics even at higher frequencies. This
belief is supported by the empirical analysis discussed in the
literature~\cite{Bouchaud_book:2000} concerning the statistical
properties of volatility proxies for intra-day data. A further perspective
would be to explore possible ways to characterize analytically the PDF
associated to the process~\eqref{eq:IGmodel} or its characteristic function.
This task requires to solve the Fokker-Planck equation for the PDF or its
equivalent version in the Fourier space, analogously to what has been done
in~\cite{DragulescuYakovenko2002} for the Heston case. Such a result would also
allow for an application of the model in the context of market risk evaluation,
possibly exploiting efficient Fourier methodologies such as those proposed
in~\cite{JStat.2010.P01005,EurPhysJB.76.157}.

\appendix
% APPENDIX: COEFFICIENTS OF THE THIRD MOMENT
\section{Coefficients of $\ang{X_t^2}$ and $\ang{X_t^3}$} \label{app:X3}

Here we report the explicit expressions of the coefficients $H^{(n)}_j(t)$
entering the expansion~\eqref{eq:XmomsExpand} of the moments of $X_t$
for the cases $n=2$ and $n=3$. They were used to plot the
analytical curves in Fig.~\ref{fig:t0scaling}.
{\scriptsize{
\begin{align*}
  H^{(2)}_0(t) &= c \,K^{(2)}_0 t \\
  H^{(2)}_1(t) &= c \,K^{(2)}_1 \left[ \frac{\exp{\left(F_1 t\right)} -1}{F_1}  \right]\\
  H^{(2)}_2(t) &= c \,K^{(2)}_2 \left[ \frac{\exp{\left(F_2 t\right)} -1}{F_2}  \right]\, ;
\end{align*}
}}
{\scriptsize{
\begin{align*}
  H^{(3)}_0(t) &= 3 \,\rho\,c^2\left\{
  \frac{t}{F_2}\left[ A_2 \frac{K^{(2)}_0}{F_1}-2K^{(3)}_0 \right]
  +2\,K^{(3)}_0 \,\left[ \frac{\exp{\left(F_2 t\right)}-1}{F_2^2} \right] \right.\\
  &\left.+A_2 \,\frac{K^{(2)}_0}{F_2-F_1} \,\left[ \frac{\exp{\left(F_2 t\right)}-1}{F_2^2}
  -\frac{\exp{\left(F_1 t\right)}-1}{F_1^2}\right]\right\} \\
  H^{(3)}_1(t) &= 3 \,\rho\,c^2\left\{\frac{1}{F_2-F_1} \left[ A_2\frac{K^{(2)}_1}{F_2-F_1}+2K^{(3)}_1\right]
  \left[\frac{\exp{\left(F_2 t\right)}-1}{F_2}\right.\right.\\
  &\left.\left.-\frac{\exp{\left(F_1 t\right)}-1}{F_1} \right]
  +A_2 \frac{K^{(2)}_1}{(F_2-F_1)F_1} \left[\frac{\exp{\left(F_1 t\right)}-1}{F_1}-t\exp{\left(F_1 t\right)} \right] \right\} \\
  H^{(3)}_2(t) &= 3 \,\rho\,c^2\left\{-A_2 \frac{K^{(2)}_2}{(F_2-F_1)^2}
  \left[\frac{\exp{\left(F_2 t\right)}-1}{F_2}-\frac{\exp{\left(F_1 t\right)}-1}{F_1}\right] \right.\\
  &\left.-\frac{1}{F_2}\left[ A_2 \frac{K^{(2)}_2}{F_2-F_1}+2 K^{(3)}_2 \right]
  \left[ \frac{\exp{\left(F_2 t\right)}-1}{F_2}-t \exp{\left(F_2 t\right)} \right]\right\}\\
  H^{(3)}_3(t) &= 6\, \rho \,c^2 \,\frac{K^{(3)}_3}{F_3-F_2}
  \,\left[\frac{\exp{\left(F_3 t\right)}-1}{F_3}-\frac{\exp{\left(F_2 t\right)}-1}{F_2} \right]\, ,
\end{align*}
}}
where the coefficients $K^{(2)}_j$ and $K^{(3)}_j$, entering the
expansion~\eqref{eq:Ymoms} of the moments of $Y_t$, read
{\scriptsize{
\begin{align*}
  K^{(2)}_0 &= \frac{A_2 A_1}{F_2 F_1} \\
  K^{(2)}_1 &= -\frac{A_2}{F_2-F_1} \left[ \mu_1(t_0)+ \frac{A_1}{F_1} \right] \\
  K^{(2)}_2 &= \mu_2(t_0) + \frac{A_2}{F_2-F_1} \left[ \mu_1(t_0)+\frac{A_1}{F_2} \right]\,;
\end{align*}
}}
{\scriptsize{
\begin{align*}
  K^{(3)}_0 &= -\frac{A_3 A_d A_1}{F_3 F_2 F_1} \\
  K^{(3)}_1 &= \frac{A_3 A_2}{(F_3-F_1)(F_2-F_1)}\,\left[
  \mu_1(t_0)+\frac{A_1}{F_1} \right] \\
  K^{(3)}_2 &= -\frac{A_3}{F_3-F_2} \left\{ \mu_2(t_0)+\frac{A_2}{F_2-F_1}
  \left[ \mu_1(t_0)+\frac{A_1}{F_2} \right]\right\} \\
  K^{(3)}_3 &= \mu_3(t_0) + \frac{A_3}{F_3-F_2}
  \left\{ \mu_2(t_0)+\frac{A_2}{F_3-F_1} \,\left[ \mu_1(t_0)+\frac{A_1}{F_3}
  \right] \right\}\,.
\end{align*}
}}

\section{Derivation of Eq.~\eqref{eq:lev3}} \label{app:leverage}

After expressing $Y_{t+\tau}$ in terms of its integral solution form $t$ to
$t+\tau$, the function $f(\tau,t;Y)$ can be rewritten in the form
{\scriptsize{
\begin{align*}
  f(\tau, t;Y) &= \ang{Y_t^2 \left( Y_t + \int_t^{t+\tau} (a Y_s + b)
  \,ds \right) \exp{\left[\sqrt{c} \Delta_t W_2(\tau)\right]}} \\
  &+ \ang{Y_t^2 \left(\sqrt{c} \int_t^{t+\tau} Y_s dW_{2,s}\right)
  \exp{\left[\sqrt{c} \Delta_t W_2(\tau)\right]}} \\
  &= \ang{\exp{\left[\sqrt{c} \Delta_t W_2(\tau)\right]}} \left[\mu_3(t) + b \tau \mu_2(t) \right] \\
  &+ a \int_t^{t+\tau} \ang{Y_t^2 Y_s \exp{\left[\sqrt{c} \Delta_t W_2(\tau)\right]}}\,ds\\
  &+ \sqrt{c} \int_t^{t+\tau} \ang{Y_t^2 Y_s \exp{\left[\sqrt{c} \Delta_t W_2(\tau)\right]} \,dW_{2,s}}~.
\end{align*}
}}
Taking into account that for $t \leq s \leq t+\tau$ we can always split
$\Delta_t W_2(\tau)$ as
{\scriptsize{
\begin{align*}
  \Delta_t W_2(\tau) &= W_{2,t+\tau} - W_{2,t} = W_{2,t+\tau} - W_{2,s} +
  W_{2,s} - W_{2,t}\\
  &= \Delta_s W_2(t+\tau-s)+\Delta_t W_2(s-t) \,,
\end{align*}
}}
the function $f(\tau,t;Y)$ becomes
{\scriptsize{
\begin{align}
  f(\tau,t;Y) &= \ang{\exp{\left[\sqrt{c} \Delta_t W_2(\tau)\right]}} \left[
  \mu_{3}(t) + b \tau \mu_2(t) \right] \nonumber\\
  &+ a \int_{0}^{\tau} \ang{Y_t^2 Y_{t+\tau'} \exp{\left[\sqrt{c}\Delta_t W_2(\tau')\right]}} \nonumber\\
  & \times \ang{\exp{\left[\sqrt{c} \Delta_{t+\tau'} W_2(\tau-\tau')\right]}}d\tau'
  \nonumber\\
  &+ \sqrt{c} \int_0^{\tau} \ang{Y_t^2 Y_{t+\tau'} \exp{\left[\sqrt{c}\Delta_t W_2(\tau')\right]}} \nonumber\\
  & \times \ang{\exp{\left[\sqrt{c} \Delta_{t+\tau'} W_2(\tau-\tau')\right]} \,dW_{2,t+\tau'}} \,,
  \label{eq:Volterra1}
\end{align}
}}
where we changed the variable of integrations to $\tau'=s-t$. Since the process
$\sqrt{c} \,\Delta_{t+\tau'} W_2(\tau-\tau')$ is Normally distributed with zero
mean and variance $c (\tau-\tau')$, and recalling the expression of the
Gaussian characteristic function, $\phi^G$, we can write
{\scriptsize{
\begin{equation*}
  \ang{\exp\left[\sqrt{c} \Delta_{t+\tau'} W_2(\tau-\tau') \right]} = \left.\phi^G (\omega) \right|_{\omega = -i}
  = \exp\left[\frac{c}{2} (\tau-\tau')\right]\,.
\end{equation*}
}}
Application of the Novikov theorem also gives
{\scriptsize{
\begin{align*} \ang{\exp{\left[\sqrt{c} \Delta_{t+\tau'} W_2(\tau-\tau')\right]}
  dW_{2,t+\tau'}} &\!=\!
  \ang{\frac{\delta\exp{\left[ \sqrt{c}\int_{t+\tau'}^{t+\tau}
  \zeta_{2,s} ds \right]}}
  {\delta \zeta_{W_2}(t+\tau')} } d\tau'\nonumber\\
  &\!=\!\sqrt{c} \,\exp\left[\frac{c}{2}(\tau-\tau')\right] \,,
  \label{eq:exp_delta_relation}
\end{align*}
}}
where we expressed the Wiener variation in terms of a Gaussian white
noise $\zeta_{2,t}$ as $dW_{2,t}=\zeta_{2,t} \,dt$. Replacing the previous
expressions in Eq.~\eqref{eq:Volterra1} we conclude that $f(\tau,t;Y)$ has
to satisfy
{\scriptsize{
\begin{multline*}
  f(\tau,t;Y) -(a+c) \int_0^\tau f(\tau',t;Y)\exp{\left[\frac{c}{2}(\tau-\tau')\right]}d\tau'=\\ 
  \exp{\left(\frac{c}{2}\tau\right)}\left[ \mu_3(t)+b \tau \mu_2(t) \right]~,
\end{multline*}
}}
which is a Volterra equation of the second kind, whose solution leads to Eq.~\eqref{eq:lev3}.

\section{Computation of $\ang{Y_t^2 Y_{t+\tau}^2 }$.} \label{app:autocorr}
With reference to the model~\eqref{eq:IGmodel}, the cross correlation
$\ang{Y_t^m Y_{t+\tau}^n}$ can be computed exactly. Provided to express
$Y_{t+\tau}^n$ as integral solution from $t$ to $t+\tau$
{\scriptsize{
\begin{equation*}
  Y_{t+\tau}^n=Y_{t}^n+\int_{t}^{t+\tau} \left[ F_n Y_s^n+A_n Y_s^{n-1}
  \right]\,ds+\int_{t}^{t+\tau}\dots dW_{2,s}
\end{equation*}
}}
it is straightforward to check that $\ang{Y_t^m Y_{t+\tau}^n}$ satisfies the
following equation
{\scriptsize{
\begin{equation}
  \frac{d}{d\tau} \ang{Y_t^m Y_{t+\tau}^n} = F_n \ang{Y_t^m Y_{t+\tau}^n} +
  A_n \ang{Y_t^m Y_{t+\tau}^{n-1}} \,,
  \label{eq:AppBcorrelRecur}
\end{equation}
}}
which is an ODE provided that the correlation $\ang{Y_t^m Y_{t+\tau}^{n-1}}$ has been
computed at the lower order $n-1$. In particular, for the case $m=n=2$, we need the
following correlation
{\scriptsize{
\begin{equation*}
  \ang{Y_t^2 Y_{t+\tau}} = \exp{\left(a\tau\right)}\mu_{3}(t)-\frac{b}{a}\,\left[1-\exp{\left(a\tau\right)}\right]\mu_{2}(t) \,;
\end{equation*}
}}
whose substitution in Eq.~\eqref{eq:AppBcorrelRecur}  provides the solution
{\scriptsize{
\begin{align*}
  \ang{Y_t^2 Y_{t+\tau}^2} &\!=\!
  \exp{\left(F_2\tau\right)} \,\mu_{4}(t)+\frac{A_2}{a-F_2}\,\left[\exp{\left(a\tau\right)}-\exp{\left(F_2\tau\right)}\right]
  \mu_{3}(t)\\
  &\!-\!\frac{b}{a}\!\left\{\frac{A_2}{F_2}\left[\exp{\left(F_2\tau\right)}-1\right]\!-\!\frac{A_2}{a-F_2}
  \left[\exp{\left(a\tau\right)}\!-\!\exp{\left(F_2\tau\right)}\right]
  \right\}\\
  &\!\times\!\mu_{2}(t)\,.
\end{align*}
}}

% ACKNOWLEDGMENTS
\begin{acknowledgments}
  We wish to acknowledge the anonymous referee for fruitful comments.  We
  sincerely thank Guido Montagna for useful discussions and comments about this
  work, and we also gratefully acknowledge the continuous support over past
  years of Oreste Nicrosini.
\end{acknowledgments}

% BIBLIOGRAPHY
%\bibliography{biblio}
%\bibliographystyle{apsrev}

\end{document}